\documentclass[aps,pra,twocolumn,showpacs,floatfix,superscriptaddress]{revtex4-1}
\usepackage{amsmath,amssymb,amsfonts,bbm,graphicx,hyperref,times}
\usepackage{color}

\renewcommand{\prl}{{Phys. Rev. Lett.} }
\renewcommand{\pra}{{Phys. Rev. A} }

\newcommand{\N}{\cal N}

\newcommand{\id}{\mathbbm{1}}

\newcommand{\gr}[1]{\boldsymbol{#1}}
\newcommand{\be}{\begin{equation}}
\newcommand{\ee}{\end{equation}}
\newcommand{\bea}{\begin{eqnarray}}
\newcommand{\eea}{\end{eqnarray}}
\newcommand{\ket}[1]{|#1\rangle}
\newcommand{\bra}[1]{\langle#1|}

\newcommand{\sig}{\gr{\sigma}}

\newcommand{\lup}[1]{\lambda^{\uparrow}_{#1}}
\newcommand{\ldn}[1]{\lambda^{\downarrow}_{#1}}

 \def\sigmaCM{\boldsymbol{\sigma}}
\newtheorem{lemma}{Lemma}
\newtheorem{proposition}{Proposition}
\newcommand{\proofend}{\hfill\fbox\\\medskip }
\newcommand{\proof}[1]{{\noindent\bf Proof } #1 $\proofend$}
\begin{document}
\title{On the optimal feedback control of linear quantum systems in the presence of thermal noise}
\author{Marco G. Genoni}
\affiliation{QOLS, Blackett Laboratory, Imperial College London, London SW7 2BW, UK}
\email{m.genoni@imperial.ac.uk}
\author{Stefano Mancini}
\affiliation{School of Science and Techonolgy, University of Camerino, I-62032 Camerino, Italy \\
and INFN, Sezione di Perugia, I-06123 Perugia, Italy}
\author{Alessio Serafini}
\affiliation{Department of Physics \& Astronomy, University College London, 
Gower Street, London WC1E 6BT, United Kingdom}

\begin{abstract}

We study the possibility of taking bosonic systems subject to quadratic Hamiltonians and a
noisy thermal environment to non-classical stationary states by feedback loops based on weak
measurements and conditioned linear driving.
We derive general analytical upper bounds for the single mode squeezing and multimode entanglement
at steady state, depending only on the Hamiltonian parameters and on the number of thermal excitations
of the bath. Our findings show that, rather surprisingly,
larger number of thermal excitations in the bath allow for larger steady-state squeezing and entanglement 
if the efficiency of the optimal continuous
measurements conditioning the feedback loop is high enough. 
We also consider the performance of feedback strategies based on homodyne detection 
and show that, at variance with the optimal measurements, it degrades with increasing temperature. 
\end{abstract}
\maketitle
\section{Introduction}\label{s:intro}
All quantum technologies hinge on establishing controlled interactions between different 
constituents of quantum systems whilst reducing unwanted interactions with an environment, which give rise to decoherence.
In dealing with environmental decoherence, two main paradigms have emerged over the last fifteen years:
one may either attempt to decouple the relevant, logical degrees of freedom from the environment by various techniques 
({\em e.g.}, decoherence free subspaces \cite{zanardi97}, error correction \cite{gottesman98}, dynamical decoupling \cite{viola99}),
and then proceed to process the quantum information coherently 
({\em e.g.}, in gate-based models of quantum computation, through unitary operations), 
or one may try to manipulate the noisy, non-unitary evolution of the system directly, 
tailoring it to suit one's aims.

The second viewpoint, which one might broadly refer to as the `dissipative' approach to quantum information processing, has a long tradition, going back 
to early proposals for reservoir engineering \cite{poyatos96}, and has recently been compounded by the design of a model for dissipative, non-unitary quantum computation \cite{verstraete11}. 
It has hence been repeatedly shown, in various contexts and settings, that
working with the environment rather than against it may lead to forms of cooperation  
whereby the environment contributes to enhance certain coherent tasks performed on the system,
often in a rather counterintuitive manner \cite{carvalho01,wang01,wang05,kraus08,ticozzi08,ticozzi09,schirmer10,stevenson11,yamamoto12,plenio99,
beige00,pastawski11,khodjasteh11,diehl11,goldstein10,krauter11,caruso10,andre11,santos11,santos11b,
bennett13,arenz13}.
Besides such enhancements, dissipative approaches typically allow for the stabilisation of target quantum resources, 
which may be a key advantage over unitary manipulation, depending on the task at hand. 

In engineering, a standard way to mould the environment to improve a system's performance, is the use of 
measurement-based feedback control. In quantum mechanics, where measurements affect the state 
of the system by inducing discontinuous jumps, 
measurement-based feedback control can be effected by monitoring part of the environment, 
which results in a weak measurement on the system, and then using the classical information 
contained in the measurement outcomes to condition subsequent manipulations of the system. 
Quantum feedback control theory blossomed over the last 10-20 years within the quantum optics, quantum control and 
quantum information communities \cite{bela,gardinerq,wisebook}, and experiments are quickly catching up 
with several successful practical demonstrations \cite{myrev}.

This paper is the account of notable cases of environmental cooperation in the 
setting of controlled dissipative dynamics in linear Gaussian systems 
\cite{wisebook,WisMilFeedback,armen02,wido05,diosiwis}.
The optimised operation of linear feedback loops to create maximal steady-state entanglement 
has been considered over the past few years, both in-loop \cite{mancini05,mancini07,seramancio}, 
and out-of-loop \cite{nurdin12}. All this body of work, 
however, is restricted to zero-temperature environments manifesting themselves through 
pure losses and no input thermal noise.  
Here, we shall consider a system of $n$ bosonic modes subject to a quadratic Hamiltonian
and to dissipation in a thermal environment with average excitation number $N$, and 
show that the maximal squeezing and entanglement achievable by continuous linear feedback control 
grows with $N$, that is with the temperature of the bath (section III).  
We will apply our results to various quadratic Hamiltonians, study quantitatively the role 
played by the efficiency of the weak measurements that condition the feedback loop, and also consider 
the problem of identifying our optimal measurement strategies, which are shown to be different from 
simple homodyne detection (section IV). The performance of feedback loops based on continuous homodyne detection 
will be studied too, and shown to degrade with increasing thermal noise.

Let us remind the reader that continuous variable squeezing and entanglement 
(the figures of merit we are considering in the present study)
hold potential for application in precision measurements \cite{xiao87,mon06,genoni11}, 
quantum information processing \cite{braunvl,gaucluster}, 
and quantum communication with continuous variable quantum systems \cite{furusawa,sqqkd}.


\section{Notation and background}\label{s:back}
We consider a system of $n$ bosonic modes described by the vector of canonical operators 
$\hat{\bf R} = (\hat{x}_1,\hat{p}_1,\ldots,\hat{x}_n,\hat{p}_n)^{\sf T}$, with commutation relations 
encoded by the anti-symmetric symplectic form $\Omega$, as per $[\hat{R}_j,\hat{R}_k]=i\Omega_{jk}$
($\hbar=1$ throughout the paper).

Being comprised of Gaussian noise, Gaussian averages, and Hamiltonian evolutions of the first (`linear driving') 
and second (`canonical', or `symplectic') order in the canonical operators, 
our dynamics will only involve Gaussian states, 
which are entirely described by first and second statistical moments of the canonical operators
\cite{ASM}. 
The second moments of a Gaussian state $\varrho$, in particular, 
will be represented by a $2n\times2n$ `covariance matrix' (CM) $\sig$:
$\sigma_{jk}={\rm Tr}(\{\hat{R}_j,\hat{R}_k\}\varrho)-2{\rm Tr}(\hat{R}_j\varrho){\rm Tr}(\hat{R}_k\varrho)$, 
which satisfies the well known 
Robertson-Schr\"odinger uncertainty relation:
\be
\sig + i \Omega \ge 0 \; . \label{heise}
\ee
This is a necessary and sufficient condition for a CM to represent a physical Gaussian state $\varrho$ \cite{Hol75}.

Let us begin by considering the most general time-independent quadratic Hamiltonian acting on the system:
$$
\hat{H}=\frac12\hat{\mathbf{R}}^\top H\hat{\mathbf{R}} \; ,
$$
where the `Hamiltonian matrix' $H$ is a generic symmetric matrix. 
We will later on modify the Hamiltonian to include a time-dependent linear term which will exert the feedback action on the system. 
The most general deterministic dynamics preserving the Gaussian character of the quantum state $\varrho$, 
taking into account the interaction with a Markovian environment, is given by a Lindblad master equation
\begin{align}
\frac{{\rm d}\varrho}{{\rm d}t} &= -i [ \hat{H},\varrho] + \sum_{j=1}^{L}\mathcal{D}[\hat{c}_j]\varrho  = \mathcal{L}_0 \varrho, \label{eq:mastereq}
\end{align}
where
\begin{align}
\mathcal{D}[O]\varrho = O\varrho O^\dagger - (O^\dagger O \varrho + \varrho O^\dag O)/2,
\end{align}
and the operators $\hat{{\bf c}}=(\hat{c}_1,\dots,\hat{c}_L)$ are linear combinations of the canonical operators, {\em i.e.} $\hat{\bf c} = \widetilde{C} \hat{\bf R}$. \\
The corresponding `free' (in that no monitoring or feedback actions have been introduced yet) dynamics of first and second moments under such conditions is described by 
\begin{align}
&\frac{{\rm d}\langle \hat{\bf R} \rangle}{{\rm d} t} = A \langle \hat{\bf R} \rangle\:, \\
&\frac{{\rm d}\sig}{{\rm d}t} = A\sig +\sig A^{\sf T}+D \:,
\end{align} 
where $A=\Omega( H + {\rm Im}[\widetilde{C}^\dag \widetilde{C}])$ and 
$D=2\Omega {\rm Re}[\widetilde{C}^\dag \widetilde{C}]\Omega^{\sf T}$.
If the system is stable, in the sense of admitting a steady state, it must be $(A+A^{\sf T})<0$, which we will assume
in what follows.\\
The matrices $A$ and $D$ are usually referred to respectively as the {\em drift} and {\em diffusion} matrix, and completely characterize the evolution of Gaussian states. 
%
We now assume to monitor continually the environment on time-scales which are much shorter than the typical system's response time, by means of weak measurements. These POVMs are usually referred to as  ``general-dyne detections''  \cite{wisebook}, encompassing all homodyne detections, both direct and resorting to ancillary modes (and hence heterodyne detection too). 
General-dyne POVMs are the most general allowing for a continuous, though stochastically fluctuating,  monitored evolution of the system. In the following, we shall distinguish between the conditional state of the system $\varrho_c$, with CM $\sig_c$  (here ``conditional'' refers to the conditioning due to the knowledge of the weak measurements' outcomes), and the time-averaged, `unconditional' state $\varrho = (1/\Delta t)\int_{t}^{t+\Delta t} \varrho_c(s){\rm d}s$, with CM $\sig$, where $\Delta t$ is an integration interval much larger than the typical time-scale of the stochastic fluctuations of the measured current \cite{note2}.
In general the evolution of the conditional state is described by the stochastic master equation (SME)
\begin{align}
{\rm d}\varrho_c = \mathcal{L}_0 \varrho_c {\rm d}t + {\rm d}{\bf z}^\dag(t) \Delta_c \hat{\bf c} \varrho_c +
\varrho_c \Delta_c \hat{\bf c}^\dag {\rm d}{\bf z}(t) \label{eq:SME}
\end{align}
where $\Delta_c \hat{O} = \hat{O} - {\rm Tr}[\varrho_c \hat{O}]$, and ${\rm d}{\bf z}=({\rm d}z_1, \dots, {\rm d}z_L )^{\sf T}$ is a 
vector of infinitesimal complex Wiener increments, with vanishing expectation values $E[{\rm d}{\bf z}]=0$. 
Each stochastic master equation, determined by the POVM describing the continuous monitoring,
is said to ``unravel'' the master equation which is obtained by averaging over the POVM's outcomes 
(in the case above, this may be done by just setting to zero all the terms where Wiener increments occur).
Hence, in the literature, a choice of the continuous monitoring is also refereed to as an ``unravelling'' 
(a terminology largely drawn from the quantum trajectories approach to open quantum systems 
\cite{gardinerq,wisebook}).

The correlations between these Wiener increments define two matrices
\begin{equation}
{\rm d}{\bf z}{\rm d}{\bf z}^\dag = \Theta {\rm d}t\:, \qquad {\rm d}{\bf z}{\rm d}{\bf z}^{\sf T}=\Upsilon {\rm d}t\:,
\end{equation}
which can be combined in a single ``unravelling'' matrix
\begin{align}
U = \frac12 \left(
\begin{array} {c c}
\Theta + {\rm Re}[\Upsilon] & {\rm Im}[\Upsilon] \\
{\rm Im}[\Upsilon] & \Theta - {\rm Re}[\Upsilon]
\end{array}
\right)\;.
\end{align}
The unravelling matrix $U$ completely characterizes the general-dyne detection performed on the environment. Notice that a proper unravelling matrix has to satisfy $U\geq 0$ and $\Upsilon^{\sf T}=\Upsilon$ \cite{diosiwis}.\\
\begin{figure}[t!]
\includegraphics[width=\columnwidth]{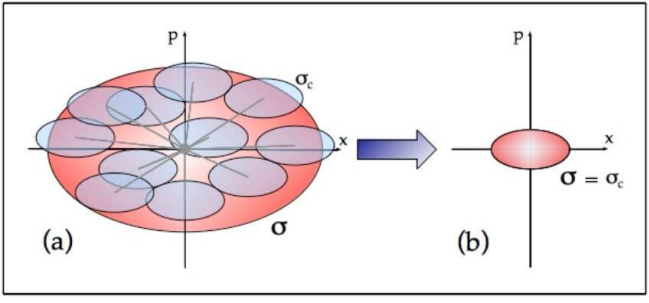}
\caption{Heuristic phase-space representation of an optimal linear feedback action. 
The unconditional state is a Gaussian average, with CM $\sig$, of conditional Gaussian states with the same CM $\sig_c$ 
and different centres in phase space (a). The optimal Markovian choice for the linear driving term, represented by gray arrows in (a) and by $\gr{u}(t)$ in the Hamiltonian, cancels the first moments of the conditional state, thus making it coincide with the 
unconditional averaged one (b). \label{Feedback}}
\end{figure}
The continuous monitoring of the output field is recorded in the general-dyne current 
\begin{align}
{\bf y}(t) = C \langle \hat{\bf R}\rangle + \frac{{\rm d}{\bf w}}{{\rm d}t} \label{eq:current}
\end{align}
where $C=(2 U)^{1/2} \bar{C}$, $\bar{C}^{\sf T}=({\rm Re}[\widetilde{C}^{\sf T},{\rm Im}[\widetilde{C}^{\sf T}])$, and ${\rm d}{\bf w}$ is a vector of real Wiener increments satisfying ${\rm d}{\bf w}{\rm d}{\bf w}^{\sf T}=\mathbbm{1}{\rm d}t$.
One can show that the dynamics of the conditional state $\varrho_c$ is Gaussian, 
with stochastic fluctuations (depending on the measured current) affecting the first moments,
but an entirely {\em deterministic} evolution for the matrix of second moments $\sig_c$
(see appendix \ref{appendixA} for details). 
This fact, as we will see, is essential to our discussion.
In fact, the white-noise fluctuations of the first moments 
are so fast that one is left with the average, unconditional evolution of the quantum state to all practical purposes. 
But, as depicted in Fig.~\ref{Feedback}, the unconditional state is just a Gaussian state resulting from the average of conditional Gaussian states with the same CM $\sig_c$ and different first moments (centres of their positions in phase space).
It is very easy to see that, under such an average, all the figures of merit we are going to consider  ({\em i.e.} entanglement and squeezing) can only decrease. 
Hence, for given general-dyne measurement, 
{\em the best case scenario for any of our figures of merit 
would be one where the fluctuations of the first moments cancel out and the average unconditional state 
coincides with the conditional state}. It turns out that such a situation can always be arranged by 
adding a linear Markovian feedback action to the Hamiltonian: 
\be
\hat{H}_f = - \hat{\bf R}^{\sf T} \Omega B {\bf y}(t) \; , \label{eq:mfb}
\ee
where ${\bf y}(t)$ is the general-dyne current and $B$ is a matrix completely determined 
by the unravelling matrix $U$ (see appendix \ref{appendixA} for the explicit expression of $B$).
Markovian feedback is therefore always optimal to our aims and we will hence restrict to it in the following.
Before proceeding, let us briefly mention that the dynamics of the averaged, unconditional second moments under 
a linear Markovian feedback action like that of Eq.~(\ref{eq:mfb})
can still be treated analytically and is of the form
${{\rm d}\sig}/{{\rm d}t}=A'\sig+\sig A'^{\sf T}+D'$ (the modified drift and diffusion 
matrices are given in the appendix \ref{appendixA}).

In view of the above, in order to optimise the steady state squeezing or entanglement, one has just to optimise the 
relevant figure of merit for the conditional state $\varrho_c$, and then apply the Markovian feedback strategy 
that ensures $\varrho=\varrho_c$ (see Fig.~\ref{Feedback}).
The optimization over the set of conditional states does not need to go into the details of the conditional dynamics but can instead be tackled by resorting to a general mathematical result: 
given drift matrix $A$ and diffusion matrix $D$, a CM $\sig_c$ is a stabilising solution of the deterministic conditional dynamics of the second moments if and only if \cite{wido05}
\be
A \sig_c+\sig_c A^{\sf T} + D \ge 0 \; . \label{dete}
\ee
In the next section we will use this last equation, together with Eq. (\ref{heise}) to derive the ultimate bounds posed by quantum mechanics on the achievable squeezing and entanglement by means of feedback strategies based on continuous general-dyne detections.
\section{Bounds on maximum achievable squeezing and entanglement}

In the following we derive analytical bounds on the Gaussian entanglement
and squeezing achievable by means of any feedback strategy 
based on general-dyne measurements and linear driving. We present our main findings 
as three lemmas  leading to two final propositions.

\begin{lemma}[Bound on smallest symplectic eigenvalue]\label{lemsympb}
The smallest partially transposed symplectic eigenvalue $\tilde{\nu}_{-}$ 
of a generic CM $\sig$ is bounded from below as follows
\be
\tilde{\nu}_{-}^2 \ge \lup{1}\lup{2} \label{sympb} \; ,
\ee
$\lup{1}$ and $\lup{2}$ being the two smallest eigenvalues of $\sig$. 
\end{lemma}
\proof{ Notice that this proof can be found in \cite{passive}. 
We will reproduce it here to make our work self-contained.

Henceforth, $\ket{v}$ will stand for a unit vector in the phase space $\Gamma$ and 
$\bra{v}$ will be its dual under the Euclidean scalar product. 
Also, given a bipartition of the modes into the `first' $l$ and the `last' $m$ modes,
let us define the matrix $T$, representing partial transposition in phase-space, as 
$T=\id_2^{\oplus l} \oplus \sigma_z^{\oplus m}$, 
$\sigma_z$ being the $z$ Pauli matrix. Hence, the partially transposed 
symplectic form is defined as $\tilde{\Omega}=T \Omega T$

The squared symplectic eigenvalue $\tilde{\nu}_{-}^{2}$ 
is the 
smallest eigenvalue of the matrix $\sig^{1/2}\tilde{\Omega}^{\sf T}\sig\tilde{\Omega}\sig^{1/2}$:
$$
\tilde{\nu}^{2}_{-} = \min_{\ket{v}} 
\bra{v}\sig^{1/2}\tilde{\Omega}^{\sf T}\sig\tilde{\Omega}\sig^{1/2}\ket{v}.
$$
For each $\ket{v}$, one can define the unit vector $\ket{w} = \tilde{\Omega}\sig^{1/2}\ket{v}/
\sqrt{\bra{v}\sig\ket{v}}$, such that $\bra{v}\sig^{1/2}\ket{w}=0$ (due to the 
antisymmetry of $\tilde{\Omega}$) and 
$$
\tilde{\nu}^{2}_{-} = \min_{\ket{v}} 
\bra{v}\sig\ket{v}\bra{w}\sig\ket{w} \ge 
\min_{\ket{v}, \ket{w}}
\bra{v}\sig\ket{v}\bra{w}\sig\ket{w} = \lup{1}\lup{2} \, .
$$
The last equality is easily verified once $\bra{v}\sig^{1/2}\ket{w}=0$
and $\sig>0$
are enforced, and completes the proof.
}


\noindent Next, the uncertainty principle entails: 

\begin{lemma}[Uncertainty relation for CMs' eigenvalues]\label{lemunc}
{Let $\{\ \lup{j} \}$ and $\{\ \ldn{j} \}$ be, respectively, the $2n$ increasingly-ordered and decreasingly-ordered
eigenvalues of an $n$-mode CM $\sig$. Then one has:
\be
\lup{j}\ldn{j} \ge 1 \quad {\rm for} \; 1\le j\le n . \label{eigunc}
\ee
}
\end{lemma}
\proof{
\noindent 
Note that the uncertainty relation (\ref{heise})
is equivalent to the two following conditions \cite{passive,pirandolo}:
\be
\sig^{1/2}\Omega^{\sf T}\sig\Omega\sig^{1/2} \ge 1 \, , \quad {\rm and} \label{heis2} \\
\quad \sig > 0 \; .
\ee
For any $\ket{v}\in\Gamma$ one can define 
$\ket{z}= \Omega\sig^{1/2}\ket{v}/\sqrt{\bra{v}\sig\ket{v}}$, 
so that the Robertson Schr\"odinger Inequality (\ref{heis2}) 
can be recast as
\be
\bra{v}\sig\ket{v} \bra{z} \sig \ket{z} \ge 1 \quad \forall\, \ket{v} \in \Gamma . \label{coso}
\ee 
We will now denote by $\ket{v_j}$ the eigenvectors corresponding to the 
increasingly ordered eigenvalues of $\sig$: $\sig\ket{v_j} = \lup{j}\ket{v_j}$.
Let us consider a vector $\ket{v}$ belonging to the subspace,
which we shall denote $\Gamma_k$, 
spanned by the $k$ smallest eigenvectors of $\sig$ $\{\ket{v_{j}}\}$, for $j\le k$.
Clearly one has $\bra{v}\sig\ket{v}\le \lup{k}$. 
The inequality (\ref{coso}) then leads to
$$
\lup{k}\bra{z} \sig \ket{z} \ge
\bra{v}\sig\ket{v} \bra{z} \sig \ket{z} \ge 1 \quad \forall\, \ket{v} \in \Gamma_k \; ,
$$
which must be satisfied by all the vectors $\ket{z}$ belonging to the $k$-dimensional linear subspace 
$\Omega\Gamma_k$ (defined as the subspace spanned by the $k$ orthogonal vectors 
$\Omega\ket{v_k}$):
$$
\lup{k}\bra{z} \sig \ket{z} \ge 1 \quad \forall \ket{z}\in \Omega\Gamma_k \; .
$$
By Poincar\'e Inequality \cite{bathia}, a vector $\ket{z}$ must exist in $\Omega\Gamma_k$ 
for which $\bra{z}\sig\ket{z} \le \ldn{k} $, such that
$\lup{k}\ldn{k}\ge 1$.
}

\noindent As an immediate corollary of Lemma 2, one obtains 
\be
\lup{1}\lup{2}\ge\frac{1}{\ldn{1}\ldn{2}}\; . \label{pollo}
\ee


\begin{lemma}[Bound on eigenvalues of steady state CMs]\label{lemssb}
Let $\sig_c$ be a conditional CM at steady state obtained under 
continuous general-dyne measurements, diffusion matrix $D$ and a 
drift matrix $A$. 
The product of the two largest eigenvalues 
$\lambda_1^\downarrow$ and $\lambda_2^\downarrow$ of
$\boldsymbol{\sigma}_c$ is bounded as follows:
\begin{align}
\lambda_1^\downarrow \lambda_2^\downarrow \leq 
\frac{(\delta_1^\downarrow  + \delta_2^\downarrow)^2}{4 \:\alpha_1^\uparrow \alpha_2^\uparrow}
\label {eigb}
\end{align}
where $\{ \alpha_j^\uparrow\}$ are the (strictly positive) eigenvalues of $(-A-A^T)$ in
increasing order, while  $\{ \delta^\downarrow\}$ are the (strictly positive) 
eigenvalues of $D$ in decreasing order. \\
\end{lemma}
\proof{
 Given the condition (\ref{dete}), 
 and given the eigenvectors of 
$\sigmaCM_c$, $|\lambda_1^\downarrow\rangle$ and $|\lambda_2^\downarrow\rangle$ corresponding to $\lambda_1^\downarrow$ and $\lambda_2^\downarrow$,
we have
\begin{align}
\lambda_1^\downarrow \langle \lambda_1^\downarrow | - (A + A^T) | \lambda_1^\downarrow\rangle &\leq \langle \lambda_1^\downarrow|D|\lambda_1^\downarrow\rangle \\
\lambda_2^\downarrow \langle \lambda_2^\downarrow | - (A + A^T) | \lambda_2^\downarrow\rangle &\leq \langle \lambda_2^\downarrow|D|\lambda_2^\downarrow\rangle 
\end{align}
By defining $\widetilde{A}=-(A+A^T)$ and multiplying the inequalities, we have
\begin{align}
\lambda_1^\downarrow \lambda_2^\downarrow 
\langle \lambda_1^\downarrow | \widetilde{A} | \lambda_1^\downarrow\rangle\langle \lambda_2^\downarrow|  \widetilde{A}  |\lambda_2^\downarrow\rangle \leq 
\langle \lambda_1^\downarrow | D | \lambda_1^\downarrow\rangle \langle \lambda_2^\downarrow | D | \lambda_2^\downarrow\rangle 
\end{align}
then
\begin{align}
\lambda_1^\downarrow \lambda_2^\downarrow &\leq \frac{\max_{\bra{v1}v2\rangle=0} \langle v_1 | D | v_1\rangle
\langle v_2 | D | v_2 \rangle}
{\min_{\bra{v1}v2\rangle=0} \langle v_1 | \widetilde{A} ) | v_1\rangle
\langle v_2 | \widetilde{A} ) | v_2 \rangle} \\
&\leq \frac{(\delta_1^\downarrow + \delta_2^\downarrow)^2}{4 \:\alpha_1^\uparrow \alpha_2^\uparrow} \label{inequa}
\end{align}
where we use
\begin{align}
\min_{\bra{v1}v2\rangle=0} &\langle v_1 | \widetilde{A}  | v_1\rangle\langle v_2 | \widetilde{A}  | v_2 \rangle 
\geq \alpha_1^\uparrow \alpha_2^\uparrow \\
\max_{\bra{v1}v2\rangle=0} &\langle v_1 | D | v_1\rangle
\langle v_2 | D | v_2 \rangle \leq \left(\frac{\delta_1^\downarrow + \delta_2^\downarrow}{2}
\right)^2
\end{align}
}

\noindent
Further, and more generally, one has: 
\begin{proposition}[Maximal unconditional squeezing]\label{squeezing}
 Let $\sig$ be the CM of a steady-state achievable 
by continuous weak general-dyne measurements and linear driving in a system of bosonic modes subject 
to a drift matrix $A$ and Gaussian white noise with a diffusion matrix $D$. 
The squeezing $\lambda^\uparrow_1$ is bounded by
\be
\lambda_1^{\uparrow} \ge \frac{\alpha^{\uparrow}_1}{\delta^{\uparrow}_{1}} \; . \label{sqb}
\ee
\end{proposition}
\proof{From Eq. (\ref{eigunc}) we obtain the relation $\lambda_1^{\uparrow} 
\geq 1/\lambda_1^{\downarrow}$, where
$\lambda_1^{\downarrow}$ $(\lambda_1^{\uparrow})$ is the largest (smallest) eigenvalue 
of a CM $\sigmaCM$. By considering a conditional CM at steady state 
and following the same line of reasoning used in 
Lemma \ref{lemssb}, we obtain the following inequality
$\lambda_1^{\downarrow} \leq \delta_1^\downarrow/\alpha_{1}^\uparrow$, which yields the inequality:
\be
\lambda_1^{\uparrow} \geq \frac{1}{\lambda_1^{\downarrow}} 
\geq  \frac{ \alpha_{1}^\uparrow}{\delta_1^\downarrow}.
\ee
As explained before, the unconditional state $\varrho$ that we obtain from our dynamics
is a statistical mixture (with Gaussian profile) of different conditional states $\varrho_{\bf r}$ having
the same CM $\sig_c$ and different first moments
${\bf r} = \langle \hat{\bf R} \rangle_c$, in formulae $\varrho=\int {\rm d}{\bf r} \: 
p({\bf r}) \varrho_{\bf r}$. As a consequence, 
the unconditional CM reads $\sig = \sig_c + \boldsymbol{\tau}$ where $\boldsymbol{\tau}>0$ is the classical covariance matrix of the 
first moments' distribution $p({\bf r})$. 
Thus the lowest eigenvalue of $\sig$ is lower bounded
by the eigenvalue of $\sig_c$ and the bound above is valid for the unconditional state.
It is worth to remember that, given an optimal CM $\sig_c$ 
which is a physical stabilising 
solution of the conditional dynamics,  the bound is tight, since 
we can always find a Markovian feedback strategy 
such that $\varrho=\varrho_{{\bf r}=0}$, that is such that  
the unconditional state has CM $\sig_c$ and zero first moments.}
\begin{proposition}[Maximal unconditional entanglement]\label{entb} 
Let $\varrho$ be the CM of a steady-state achievable by continuous 
weak general-dyne measurements and linear driving in a system of bosonic modes subject to 
a drift matrix $A$ and Gaussian white noise with a diffusion matrix $D$. 
The logarithmic negativity $E_{\N}(\varrho)$ \cite{vidal02} of any $1$ versus $(n-1)$ modes 
or bisymmetric bipartition of $\varrho$ is bounded by
\be
E_{\N}(\varrho) \le \max\left[0,
\log_2 \left(\frac{\delta^{\downarrow}_1+\delta^{\downarrow}_2}{2\sqrt{\alpha^{\uparrow}_1\alpha^{\uparrow}_2}}\right)\right] \; . \label{entb}
\ee
\end{proposition}
\proof{
 The chain of Inequalities (\ref{sympb}), (\ref{pollo}) and (\ref{eigb}) leads to 
\be
\tilde{\nu}_-^2 \geq \frac{4 \:\alpha_1^\uparrow \alpha_2^\uparrow}{(\delta_1^\downarrow + \delta_2^\downarrow)^2}
\; , \label{feedbound}
\ee
which, in turn, constrains the maximal logarithmic negativity 
achievable for states $\varrho_{\bf r}$ conditioned by Gaussian measurements
having a CM $\sig_c$. 
In fact, by using the formula $E_{\N}=\max[0,-\log(\tilde{\nu}_{-})]$ ,
we obtain,
\begin{align}
E_{\N}(\varrho_{\bf r}) \le \max\left[0,
\log_2 \left(\frac{\delta^{\downarrow}_1+\delta^{\downarrow}_2}{2\sqrt{\alpha^{\uparrow}_1\alpha^{\uparrow}_2}}\right)\right] \; . \label{entb}
\end{align}
On the other hand the unconditional (Gaussian)
state reads $\varrho = \int d{\bf r} \: 
p({\bf r}) \varrho_{\bf r} $; this implies that $\varrho$ can be obtained from the 
Gaussian state $\varrho_{{\bf r}=0}$ (having CM $\sig_c$ and vanishing first moments) 
by local operations and classical communication alone, because first moments can be arbitrarily adjusted by 
local unitary operations. 
Since the log-negativity is an entanglement monotone \cite{plenioln}, we have
$E_{\N}(\varrho) \leq E_{\N}(\varrho_{\bf r}),$ that is the 
bound above is valid also for the unconditional state and can be achieved 
by means of optimal Markovian feedback.}

\subsection{Remarks on the bounds}
A noticeable feature of both our bounds is that they increase (somewhat loosely, we will refer to 
$\lambda_1^{\downarrow}$ getting smaller as an `increase' in the squeezing) if the largest eigenvalues of 
the diffusion matrix $D$ increase, which characterises a noisier environment. 
As we will see in the following section, if one considers a simple thermal environment, 
the diffusion matrix reads $D=\bigoplus_{j=1}^n (1+2N_j)\mathbbm{1}_2$, 
and thus
$\delta^{\downarrow}_1=\delta^{\downarrow}_2=1+2N^{\downarrow}_1$, where $N^{\downarrow}_1$ 
is the largest number of thermal excitations in an environmental degree of freedom. 
As already proven in \cite{seramancio} in the special case of pure losses, our bounds are actually tight for several important dynamics, where they represent the actual maximal values achievable. 
Exact general conditions for the tightness of the bounds are presented in Appendix \ref{appendix}, 
while specific important instances are treated in section IV.  

\begin{figure}[t!]
\begin{center}
\includegraphics[width=\columnwidth]{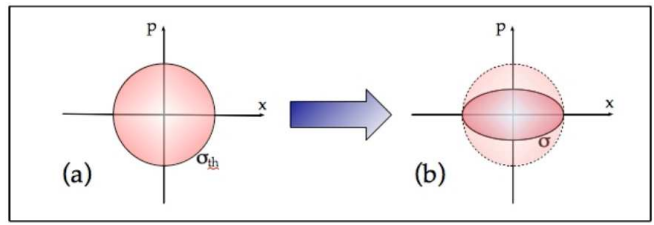}
\caption{Heuristic phase space picture of the noise-enhancement of the optimal feedback action. 
The feedback squashes the thermally broadened unconditional steady state CM $\sig_{\rm th}$, turning it into the squeezed CM $\sig$.
By Heisenberg principle, the squashing is limited by the inverse of the thermal uncertainty in the orthogonal quadrature,
which increases with increasing noise.
 \label{Unconditional}}
\end{center}
\end{figure}

Hence, our findings show that the maximal achievable entanglement increases with the temperature of the bath. 
This apparently counterintuitive behaviour can be illustrated and understood by 
considering the feedback action on the squeezing of the {\em unconditional} state 
of a free single bosonic mode \cite{note4}. 
As we shall see, in this case the optimal procedure to obtain squeezing consists in 
monitoring the environment through a specific general-dyne POVM -- also known as a specific ``unravelling'', 
along a given phase space direction (in the sense that the average of the general-dyne current coincides with the 
expectation value of the quadrature along that direction in phase space), 
and then in systematically driving the expectation value of the monitored quadrature to zero. 
As illustrated in Fig.~\ref{Unconditional}, this produces an uncertainty contraction 
for that quadrature, while the conjugate, orthogonal quadrature is entirely unaffected. 
Hence, by the Heisenberg principle, the achievable squeezing is 
ultimately limited by the inverse of the uncertainty in the orthogonal quadrature,
which clearly increases with the available thermal energy of the bath. 
In a sense, this is a case of reservoir engineering 
where the effect of the bath is `squashed' \cite{squash}, rather than squeezed, by means of continuous measurements.

If one is interested in optimal squeezing, this thermal enhancement can be obtained by measuring 
and acting locally on a single quadrature, while the generation of optimal entanglement will generally require nonlocal measurements. 
It should however be noted here that linear feedback does allow for an increase in steady-state unconditioned entanglement even with local measurements, if the Hamiltonian couplings 
between the modes are strong enough \cite{seramancio}.
It should also be noted that, whenever the bound is achievable, the optimal steady-state is pure, because the saturation of the 
uncertainty relation (\ref{heise}) is implied. In such cases, the optimal feedback strategy not only maximises a figure of merit but also stabilises a pure state, regardless of how noisy the environment may be.

We should also note that the squeezing and entanglement optimised 
in our analysis are {\em in-loop}, rather than out of loop, resources. 
Depending on the specifics of the considered set-up, in practice one might get around this problem 
by: (i) turning off the control such that the resources (squeezing and entanglement) of the system will be 
transferred to output fields on short enough time scales (see, {\em e.g.}, \cite{tomcanali}); 
(ii) including the additional systems that have to exploit the quantum resources in the feedback loop, 
as was suggested for example in \cite{wisemanInLoop}. For a treatment focusing on the 
out-of-loop entanglement transferable to travelling modes, encompassing the effect of delays 
and losses at zero temperature, see \cite{nurdin12}.


\section{Optimal and homodyne-based feedback strategies}
In this section we will evaluate the bounds for some
cases with direct experimental relevance, and contrast optimal performances with 
what can be obtained with readily-available homodyne detection. 

We will focus on the case of a finite temperature Markovian
environment, in which case Eq. (\ref{eq:mastereq}) can be rewritten as
\begin{align}
\frac{{\rm d}\varrho}{{\rm d}t} &=\mathcal{L}_{\rm th}\varrho \\
&= -i[\hat{H},\varrho] + \kappa\sum_{j=1}^{n} \left[ (N_j+1) \mathcal{D}[\hat{a}_j]\varrho + N_j \mathcal{D}[\hat{a}^\dagger] \varrho \right] 
\end{align}
where $\kappa$ is the loss rate of the system and $N_j$ represents the number of the thermal
excitations in the bath of mode $j$ \cite{note1}. The drift and diffusion matrices then read
\begin{align}
A_{\rm th}&=(\Omega H - \kappa \mathbbm{1}_{2n})/2, \\
D_{\rm th}&=\bigoplus_{j=1}^n (1+2N_j)\mathbbm{1}_2.
\end{align}
In the following we will calculate the bounds on squeezing and entanglement and
present both the stochastic
master equation corresponding to the optimal strategies saturating
the bounds, as well as the ones based on homodyne linear feedback.

\subsection{Free System}
Let us start with the simple case where no Hamiltonian is present ($H=0$,
which in practice corresponds to considering a system in the rotating frame, and to having all the 
measurements' phase references rotate accordingly). 
Henceforth, we will always set $\delta^{\downarrow}_1=\delta^{\downarrow}_2=1+2N^{\downarrow}_1$ 
(phase-insensitive thermal noise). 
{Without any feedback action, the steady-state clearly corresponds to
a thermal state without squeezing nor entanglement}.
On the other hand, the bounds on the squeezing and logarithmic negativity 
achievable via feedback read, respectively \cite{kappa}, 
\begin{align}
\lambda_1^{\uparrow} &\ge 1/(1+2N^{\downarrow}_1) \:, \\
E_{\N} &\le \log_2 \left(1+2N^{\downarrow}_1\right)\:.
\end{align}
As regards single-mode squeezing, one can show that the bound is achievable if one implements a continuous measurement on the environment described by the following stochastic master equation
\begin{align}
{\rm d}\varrho_c &= \mathcal{L}_{\rm th}\varrho_c \: {\rm d}t + \sqrt{N^{\downarrow}_1+1} \mathcal{H}[\hat{a}e^{i\phi}]\varrho_c \: {\rm d}w_1 \: + \nonumber \\
&\qquad + \sqrt{N^{\downarrow}_1}\mathcal{H}[\hat{a}^\dag e^{-i\phi}] \varrho_c \: {\rm d}w_2 \:, \label{smeofs}
\end{align}
where $\hat{a}$ represents the mode that we want to squeeze, interacting with the bath having $N^{\downarrow}_1$ thermal photons, ${\rm d}w_j$ are Wiener increments that satisfy ${\rm d}w_j {\rm d}w_k = \delta_{jk}$, and
\begin{align}
\mathcal{H}[\hat{O}]\varrho = \hat{O}\varrho + \varrho \hat{O}^\dag - {\rm Tr}[\varrho (\hat{O}+\hat{O}^\dag)] \:.
\end{align} 
The strategy is based on a POVM parametrized by two real continuous values 
with respective currents both proportional to the average value of the quadrature $\langle \hat{x}_\phi \rangle$ that we intend to squeeze. The corresponding Markovian feedback strategy is straightforwardly based on driving the orthogonal quadrature by means of these currents. The practical realization of such a a continuous measurement is a different problem that should be addressed separately. \\
One may wonder what the result is if a simple continuous homodyne measurement of the 
bath is performed, described by the SME
\begin{align}
{\rm d}\varrho_c &= \mathcal{L}_{\rm th}\varrho_c \: {\rm d}t + \frac{1}{\sqrt{2N^{\downarrow}_1+1}} 
\left\{ (N^{\downarrow}_1+1) \mathcal{H}[\hat{a}e^{i\phi}] \nonumber \right. \\
& \qquad \left. - N^{\downarrow}_1\mathcal{H}[\hat{a}^\dag e^{-i\phi}] \right\} \varrho_c \: {\rm d}w \:, \label{smehfs}
\end{align}
where a single real Wiener increment ${\rm d}w$ is present. It is easy to prove that in this case the steady state 
covariance matrix of the conditional state (that one obtains unconditionally by means of Markovian feedback) is ${\boldsymbol \sigma}=D_{\rm th}$. No squeezing can be produced and the feedback action does not bear any effect on the 
steady state. Direct comparison of Eqs.~(\ref{smeofs}) and (\ref{smehfs}) show that homodyne detection coincides 
with the optimal strategy at zero temperature, where $N_1^{\downarrow}=0$ (which is however uninteresting since the 
steady state is just the vacuum in such a case).

{As for the entanglement, we can show that in the two-mode ($n=2$) case,
if the two baths have the same temperature $N$, the bound 
can be saturated. One of the optimal unravellings is described by the SME
\begin{widetext}
\begin{align}
{\rm d}\varrho_c &= \mathcal{L}_{\rm th}\varrho_c \: {\rm d}t + \sqrt{\frac{N+1}{2}} \mathcal{H}[\hat{a}+\hat{b}]\varrho_c \: {\rm d}w_1 \: + \sqrt{\frac{N}{2}}\mathcal{H}[\hat{a}^\dag+\hat{b}^\dag] \varrho_c \: {\rm d}w_2 + \sqrt{\frac{N+1}{2}} \mathcal{H}[i(\hat{b}-\hat{a})]\varrho_c \: {\rm d}w_3 \: + \nonumber \\
&\qquad + \sqrt{\frac{N}{2}}\mathcal{H}[i(\hat{a}^\dag-\hat{b}^\dag )] \varrho_c \: {\rm d}w_4 \:, \label{eq:entopt}
\end{align}
\end{widetext}
where, as usual, ${\rm d}w_j {\rm d}w_k = \delta_{jk}$. This corresponds to a non-local strategy with four currents, 
such that the average of two of the four components of the current vector ${\bf y}(t)$ are proportional to the expectation value 
$\langle \hat{x}_a - \hat{x}_b\rangle$, and the remaining two correspond to $\langle \hat{p}_1 + \hat{p}_2 \rangle$. 
The entangled steady state can be obtained unconditionally by driving respectively the quadratures  
$(\hat{p}_1 - \hat{p}_2)$ and $( \hat{x}_a + \hat{x}_b)$. 

One can also analytically include an efficiency parameter $\eta$ for the continuous measurements performed,
with $0\le\eta\le 1$.  
This is incorporated by assuming the loss of a portion $(1-\eta)$ of the amplitude hitting each detector (equivalent 
to the action of a beam splitter with transmittivity $\eta$ before the detectors).
The logarithmic negativity achieved for efficiency $\eta$ is given by
$E_{\N}=\log_2\left(1+2 {N}^{\downarrow}_1\right)
-\log_2\left(1+4 {N}^{\downarrow}_1(1-\eta)+4{N}^{\downarrow}_1(1-\eta)\right)$. 
By inspecting this equation one observes that, for a given temperature
$N$, one can define a threshold value 
$\eta_{\rm th}=\frac{1+2N}{2(1+N)}$ 
such that entanglement is obtained only for efficiencies $\eta > \eta_{\rm th}$.
We notice that $\eta_{\rm th}$ is always greater than $1/2$ and monotonically
increases with temperature towards the maximum value corresponding to 
a perfect measurement. One could for example consider the practical 
consequences, in systems where very few thermal excitations are the dominant
source of noise, as could be the case at terahertz frequencies in solid-state and
optical systems at room temperature \cite{teraphonons, teraphotons}: 
if $N\simeq 1$ (corresponding to about 4 THz), then the threshold value is
$\eta_{\rm th}=0.75$. The optimal efficiency raises very quickly to $1$, being already
$\eta \simeq 0.9$ for $N \simeq 5$ (around $1$ THz).


It is possible to contrast these findings with the effect of a (non-local) continuous homodyne detection of the quadratures  
$\hat{x}_a - \hat{x}_b$, and $\hat{p}_1 + \hat{p}_2 $, described by the SME
\begin{widetext}
\begin{align}
{\rm d}\varrho_c &= \mathcal{L}_{\rm th}\varrho_c \: {\rm d}t + \frac{1}{\sqrt{2N+1}} 
\left\{ \frac{N+1}{\sqrt{2}} \mathcal{H}[\hat{a}+\hat{b}]  - \frac{N}{\sqrt{2}}\mathcal{H}[\hat{a}^\dag +\hat{b}^\dag] \right\} \varrho_c \: {\rm d}w_1 + \nonumber \\
&\qquad + \frac{1}{\sqrt{2N+1}} 
\left\{ \frac{N+1}{\sqrt{2}} \mathcal{H}[i(\hat{b}-\hat{a})]  - \frac{N}{\sqrt{2}}\mathcal{H}[i(\hat{a}^\dag -\hat{b}^\dag)] \right\} \varrho_c \: {\rm d}w_2 \:, \label{eq:homo2}
\end{align}
\end{widetext}
accompanied by the linear driving of the conjugated quadrature. As we saw above for a single homodyne detection, 
no action on the steady state is obtained in such a case, being the corresponding covariance matrix ${\boldsymbol \sigma}=D_{\rm th}$.} Like for the single-mode case, direct comparison of Eqs.~(\ref{eq:entopt}) and (\ref{eq:homo2}) show that homodyne detection is optimal for $N_1^{\downarrow}=0$ (which is however uninteresting since the 
steady state is just the vacuum in such a case).

Also note that, in the most general case, when $N^{\downarrow}_1\neq N^{\downarrow}_2$, 
the bound cannot always be saturated. 
However, we were able to find an unravelling similar to the one in Eq. (\ref{eq:entopt}), where 
${N}$ has to be replaced by ${N}^{\downarrow}_2$ and
the steady state is a pure two-mode squeezed state  with logarithmic negativity 
$E_{\N}=\log_2\left(1+2{N}^{\downarrow}_2 \right)$, which still highlights a thermally-enhanced performance. 
\subsection{Parameteric Hamiltonians}
We now move on to consider the case of degenerate parametric down conversion
which can be described, in interaction picture, by the quadratic Hamiltonian 
$$
\hat{H}=\chi(\hat{x}_1\hat{p}_2+\hat{p}_1\hat{x}_2)
$$
between two modes at the same frequency \cite{reid89},
such that the average number of thermal excitations in the two modes are the same and set equal to $N$, yielding 
\begin{align}
A&=-\frac{\mathbbm{1}_4}{2} + \frac{\chi}{2} \left(
\begin{array}{c | c}
0 & \sigma_z \\
\hline
\sigma_z & 0
\end{array}
\right)\:, \\
D_{\rm th}&= (1+2N) \id_4 \; ,\:
\end{align}
where $\sigma_z$ is the Pauli $z$ matrix.
We shall impose stability by bounding the interaction strength: $\chi<1/2$.
This set of dynamical parameters allows for the perfect saturation of the bound on the entanglement, and thus for the analytical optimisation of the achievable logarithmic negativity, which is given by $$
E_{\N}\leq\log_2(1+2N)-\log_2(1-2\chi)\:,
$$
to be compared with the free steady state value 
$$E_{\N}^{(0)} =\log_2(1+2\chi)-\log_2(1+2N) \:,
$$
that would be obtained in the 
absence of monitoring and feedback action. 
{This is possibly the most apparent example of noise-enhanced performance in our study: while the 
free steady state logarithmic negativity decreases with the temperature, 
as one would expect, its optimised counterpart increases with $N$}.
Closed-loop general-dyne control is in principle capable of 
retrieving information from the output channel and turning the phase insensitive thermal energy into correlations between the modes. The optimal feedback strategy 
can be determined in this case as well: it
corresponds to a continuous measurement described by the SME in Eq. (\ref{eq:entopt}),
followed by driving the quadratures $(\hat{x}_1+\hat{x}_2)$ and $(\hat{p}_1-\hat{p}_2)$ with the currents obtained by monitoring the conjugated squeezed 
quadratures $(\hat{p}_1+\hat{p}_2)$ and $(\hat{x}_1-\hat{x}_2)$, respectively.
In this case too a perfect measurement is required and 
one should hence consider the effect of the efficiency $\eta$ on the achievable optimal entanglement. 
The conditions on the measurement 
efficiency for the feedback loop to be able to improve the generation of entanglement are rather strict,  and become steeper and steeper as the noise increases. 
For $\chi=0.3$ and $N=1$, where the steady state in absence of feedback is unentangled, 
$\eta\ge0.8$ is needed to generate any entanglement between the two set of modes. 
This threshold increases to $0.92$ for $N=2.5$ and to $0.98$ for $N=10$. These are hence the typical values 
of excitations where linear feedback control might really make a difference in the generation of 
pure entangled states of continuous variable systems. 

Once again, we can contrast this result with a feedback strategy based on the weak, continuous homodyne detection 
of the quadratures $(\hat{p}_1+\hat{p}_2)$ and $(\hat{x}_1-\hat{x}_2)$, described by Eq. (\ref{eq:homo2}). In the case of zero temperature, the two approaches coincide, as already shown in \cite{seramancio}. 
For non zero temperature, the entanglement achievable by homodyne detection of the bath is instead equal to
\begin{align}
E_{\N}(\varrho) = \max\left\{ 0 , -\log_2\left[(1+2N)(1-2\chi)\right] \right\} \: .
\end{align}
The steady state entanglement attainable by homodyne detection decreases with increasing temperature. 
Moreover one can define a threshold value 
$$
\chi_t(N) = \frac{N}{1+2 N},
$$
such that entanglement can be obtained only for values of the coupling constant $\chi > \chi_t(N)$.
\section{Conclusions} 
In this paper, we have derived bounds on single-mode squeezing and two-mode Gaussian 
entanglement achievable by means of continuous measurement and feedback on a bosonic quantum system interacting with a Markovian thermal environment.
We have shown that these bounds improve by increasing the temperature
of the bath, and derived optimal continuous unravelling, in terms of stochastic master equations, 
to attain such bounds in the cases of parametric interactions and free systems. 
We have also shown that, by restricting to homodyne continuous measurements, the expected 
dependence on temperature of the achievable figures of merit are recovered. Optimal performances 
correspond to homodyne measurements only in the zero-temperature limit.

The implementation of the optimal unravellings introduced here, 
which are able not only to stave off the effect of thermal noise and achieve pure 
steady states \cite{smith02,sayrin12}, but also to, in a sense, convert the thermal energy of the bath 
into enhanced squeezed or entangled resources, will be the object of future inquiry. 

\section{Acknowledgments} The authors thank Andrew Doherty, Matteo Paris and Howard Wiseman for useful and interesting discussions. MGG acknowledges support from UK EPSRC (EP/I026436/1).

\appendix
\section{Linear quantum systems and optimal unravellings} \label{appendixA}
In this appendix we will provide the reader with details about the evolution of
Gaussian quantum states under continuous
quantum measurements and linear feedback.\\
We will start by considering the SME in Eq. (\ref{eq:SME}) 
describing continuous general-dyne measurements.
For the conditional
state, we obtain a diffusive equation with a stochastic component for the first moments 
$\langle \hat{\bf R} \rangle_c$, and a 
deterministic equation for the CM $\sig_c$. In formulae:
\begin{align}
{\rm d} \langle \hat{\bf R} \rangle_c  &= A \langle \hat{\bf R}\rangle_c {{\rm d}t}+ (\sig_c C^{\sf T} +
\Gamma^{\sf T}) {\rm d}{\bf w}\, ,  \label{eq:firstmom} \\
\frac {{\rm d} \sig_c} {{\rm d}t} =& A \sig_c + \sig_c A^{\sf T} + D - 
(\sig_c C^{\sf T} + \Gamma^{\sf T})(C \sig_c + \Gamma)\, , \label{eq:condsig}
\end{align}
where ${\rm d}{\bf w}$ is a vector
of real Wiener increments satisfying ${\rm d}{\bf w}{\rm d}{\bf w}^{\sf T} = \id_{2n}\,{\rm d}t$ \cite{gardiner},
 $\Gamma = (2 U)^{1/2} S \bar{C} \Omega$ and 
$$
S=\left(
\begin{array} { c c }
0 & \mathbbm{1}_{\tiny n} \\
-\mathbbm{1}_{\tiny n} & 0
\end{array}
\right).
$$
\\
Then, we consider the addition of a linear time-dependent term 
to the Hamiltonian:
\be
\hat{H}_f = - \hat{\bf R}^{\sf T} \Omega B {\bf y}(t) \; , \label{fb} 
\ee
where ${\bf y}(t)$ is the current obtained from the continuous 
measurement in Eq.~(\ref{eq:current}) and the matrix
$B$ defines the Markovian feedback action exerted on the system.
In this case, the evolution equation for the unconditional state 
covariance matrix $\sig$ is still of the form $(d \sig)/dt = A' \sig + \sig A'^{\sf T} + D'$, 
where
\begin{align}
A' &= A + BC \; ,\\ 
D' &= D - C^{\sf T} B^{\sf T} - BC +2B B^{\sf T},
\end{align}
and $C=2 (U)^{1/2} \bar{C}$ . \\
A CM $\sig_c$ is a physical stabilising solution of 
the conditional dynamics if it satisfies the two following conditions [see Eq.~(\ref{eq:condsig})
and notice that the second term on the right-hand side is always positive]
\begin{align}
\sig_c+ i \Omega &\geq 0 \qquad {\rm (physicality}\:\:{\rm condition)\, ,} \label{heis} \\
A \sig_c + \sig_c A^{\sf T} + D &\geq 0 \qquad {\rm (stabilising}\:\:{\rm condition) \, .} 
\label{eq:stab}
\end{align}
As derived by Wiseman and Doherty \cite{wido05}, given a stabilising CM $\sig_c$, an
optimal unravelling $U_{\rm opt}$ such that $\sig_c$ can be obtained at steady-state,
always exists. In particular a (not necessarily unique) optimal unravelling $U_{\rm opt}$ 
can be obtained by solving the equation
\begin{align}
2 E^{\sf T} U E = D + A \sig_c + \sig_c A^{\sf T} \; , \label{eq:optun}
\end{align}
where $E= \bar{C} \sig_c + S \bar{C} \Omega$. 
The Hamiltonian term in Eq. (\ref{fb}) is then chosen so as to cancel out the first
moments and make the average unconditional state coincide with the conditional state. 
It can be shown that the matrix 
$B_{\rm opt}$ achieving this, for a given steady-state CM $\sig_c$, reads
\begin{align}
B_{\rm opt} = - \sig_c C^{\sf T} -\Gamma^{\sf T}.
\end{align}

\section{Necessary conditions for the tightness of the bounds}\label{appendix}
In this appendix, by considering how our bounds were derived, and working backward, we will determine sharp conditions 
on the matrices $A$ and $D$, for the bounds to be achievable.
In order to express such conditions, let us define the eigenvectors $\ket{\alpha^{\uparrow}_j}$ and 
$\ket{\delta^{\downarrow}_j}$ associated, respectively, to the $j-$th smallest eigenvalue of 
$\tilde{A}=-A-A^{\sf T}$ and $j$-th largest eigenvalue of $D$.
This leads to the following two additional propositions: 
\begin{proposition}[Conditions for maximal squeezing] 
A continuously measured and linearly driven Gaussian system is able to saturate the bound (\ref{sqb}) if and only if 
\be
\ket{\alpha^{\uparrow}_1} = \ket{\delta^{\downarrow}_1} \; . \label{condsq}
\ee
\end{proposition}
\proof{The inequality $\lambda^{\downarrow}_1\le\delta^{\downarrow}_1/\alpha_1^{\uparrow}$ (the analogous of (\ref{inequa}) for the squeezing case) is only saturated if the eigenvector 
associated to the largest eigenvalue of $\sig$ coincides with 
$\ket{\alpha^{\uparrow}_1}$ and $\ket{\delta^{\downarrow}_1}$, hence our condition (\ref{condsq}),
in that it is always possible to construct a physical $\sig$ with largest eigenvalue along a particular direction.}
\begin{proposition}[Conditions for maximal entanglement] 
A continuously measured and linearly driven Gaussian system is able to saturate the bound (\ref{entb}) if and only if 
the following relationships are satisfied
\bea
\ket{\alpha^{\uparrow}_1} &=& \frac{\ket{\delta^{\downarrow}_1} \mp \ket{\delta^{\downarrow}_2}}{\sqrt{2}} \; ,  \label{condent1} \\
\ket{\alpha^{\uparrow}_2} &=& \frac{\ket{\delta^{\downarrow}_1} \pm \ket{\delta^{\downarrow}_2}}{\sqrt{2}} \; , \label{condent2}\\
\ket{\alpha^{\uparrow}_2} &=& \Omega^{\sf T}\tilde{\Omega}\Omega \ket{\alpha^{\uparrow}_1} \; , \label{condent3} \\
\bra{\alpha^{\uparrow}_1} T \ket{\alpha^{\uparrow}_1} &=& 0  \label{condent4}
\eea
(where $\mp$ and $\pm$ mean that if Eq.~(\ref{condent1}) has a minus sign then Eq.~(\ref{condent2}) has a plus, and viceversa, 
and that either choice is a valid condition).
\end{proposition}
\proof{Eqs.~(\ref{condent1}) and (\ref{condent2}) are necessary for the saturation of Inequality (\ref{inequa}), 
along with the choices $\ket{\lambda^{\downarrow}_1}=\ket{\alpha^{\uparrow}_1}$ and 
$\ket{\lambda^{\downarrow}_2}=\ket{\alpha^{\uparrow}_2}$. 
Then, inspection of the proof of Lemma 2 for $k=1$, $k=2$ and, by induction, for any $k$, 
reveals that the condition $\lambda^{\downarrow}_k\lambda^{\uparrow}_{k}=1$ is saturated if and only if 
$\ket{\lambda^{\uparrow}_k}=\Omega\ket{\lambda^{\downarrow}_{k}}$ (where the eigenvectors 
associated to $\lambda^{\downarrow}_k$ and $\lambda^{\uparrow}_k$ 
have been denoted with $\ket{\lambda^{\downarrow}_k}$ and $\ket{\lambda^{\uparrow}_k}$). 
Now, in order to saturate the bound, this additional condition can only be imposed if the two eigenvectors 
$\ket{\lambda^{\downarrow}_{k}}$,
already determined by (\ref{condent1}) and (\ref{condent2}), are orthogonal to 
$\Omega\ket{\lambda^{\downarrow}_{k}}$ (so that the latter can also be eigenvectors of $\sig$), that is
\be
\bra{\lambda^{\downarrow}_{2}}\Omega \ket{\lambda^{\downarrow}_{1}} = \bra{\alpha^{\downarrow}_{2}}\Omega \ket{\alpha^{\downarrow}_{1}} = 0 \; . \label{auscond} 
\ee
Further, inspection of Lemma 1 shows that, for Inequality (\ref{sympb}) to be saturated, it must be 
$\ket{\lambda^{\uparrow}_{2}}=\tilde{\Omega}\ket{\lambda^{\uparrow}_{1}}$ which, by the 
conditions $\ket{\lambda^{\downarrow}_1}=\ket{\alpha^{\uparrow}_1}$, 
$\ket{\lambda^{\downarrow}_2}=\ket{\alpha^{\uparrow}_2}$ and 
$\ket{\lambda^{\uparrow}_{k}}=\Omega\ket{\lambda^{\downarrow}_{k}}$ imposed at previous steps, becomes 
$\ket{\alpha^{\uparrow}_2} = \Omega^{\sf T}\tilde{\Omega}\Omega \ket{\alpha^{\uparrow}_1}$, which proves 
condition (\ref{condent3}).
Finally, by inserting Eq.~(\ref{condent3}) into Eq.~(\ref{auscond}), and noting that 
$\Omega^{\sf T}\tilde{\Omega}\Omega \Omega = -T$, one can recast condition (\ref{auscond}) 
in terms of $\ket{\alpha^{\uparrow}_1}$ alone as condition (\ref{condent4}).}



\begin{thebibliography}{99}

\bibitem{zanardi97}P. Zanardi and M. Rasetti, Phys. Rev. Lett. {\bf 79}, 3306 (1997).

\bibitem{gottesman98} D. Gottesman, Phys. Rev. A {\bf 57}, 127 (1998).

\bibitem{viola99} L. Viola, E. Knill, and S. Lloyd, Phys. Rev. Lett. {\bf 82}, 2417 (1999). 

\bibitem{poyatos96}J. F. Poyatos, J. I. Cirac, and P. Zoller, Phys. Rev. Lett. {\bf 77}, 4728 (1996).

\bibitem{verstraete11} F. Verstraete, M. M. Wolf, and J. I. Cirac, Nature Phys. {\bf 5}, 633 (2009).

\bibitem{carvalho01}A. R. R. Carvalho, P. Milman, R. L. de Matos Filho, and L. Davidovich, Phys. Rev. Lett. {\bf 86}, 4988 (2001).

\bibitem{wang01}J. Wang and H. M. Wiseman, Phys. Rev. A {\bf 64}, 063810 (2001).

\bibitem{wang05}J. Wang, H. M. Wiseman, and G. J. Milburn, Phys. Rev. A {\bf 71}, 042309 (2005).

\bibitem{kraus08} B. Kraus, H. P. Buchler, S. Diehl, A. Kantian, A. Micheli, and P. Zoller, Phys. Rev. A {\bf 78}, 042307 (2008).

\bibitem{ticozzi08}F. Ticozzi and L. Viola, IEEE Trans. on Aut. Control {\bf 53}, 2048 (2008).
 
\bibitem{ticozzi09}F. Ticozzi, S. G. Schirmer, and X. Wang, IEEE Trans. on Aut. Control {\bf 55}, 2901 (2010).

\bibitem{schirmer10}S. G. Schirmer and X. Wang, Phys. Rev. A {\bf 81}, 062306 (2010).

\bibitem{stevenson11}R. N. Stevenson, J. J. Hope, and A. R. R. Carvalho, Phys. Rev. A {\bf 84}, 022332 (2011).

\bibitem{yamamoto12}K. Koga and N. Yamamoto, Phys. Rev. A {\bf 85}, 022103 (2012).

\bibitem{plenio99}M. B. Plenio, S. F. Huelga, A. Beige, and P. L. Knight, Phys. Rev. A {\bf 59}, 2468 (1999). 

\bibitem{beige00}A. Beige, D. Braun, B. Tregenna, and P. L. Knight, Phys. Rev. Lett. {\bf 85}, 1762 (2000).

\bibitem{pastawski11}F. Pastawski, L. Clemente, and J. I. Cirac, Phys. Rev. A {\bf 83}, 012304 (2011).

\bibitem{khodjasteh11}	K. Khodjasteh, V. V. Dobrovitski, and L. Viola, Phys. Rev. A {\bf 84}, 022336 (2011).

\bibitem{diehl11}S. Diehl, E. Rico, M. A. Baranov, and P. Zoller, Nature Phys. {\bf 7}, 971 (2011). 

\bibitem{goldstein10}G. Goldstein, P. Cappellaro, J. R. Maze, J. S. Hodges, L. Jiang, A. S. S\"orensen, and M. D. Lukin, Phys. Rev. Lett. {\bf 106}, 140502 (2011).

\bibitem{krauter11}H Krauter, C. A. Muschik, K. Jensen, W. Wasilewski, J. M. Petersen, J. I. Cirac, and E. S. Polzik, Phys. Rev. Lett. {\bf 107}, 080503 (2011). 

\bibitem{caruso10}F. Caruso, S. F. Huelga, and M. B. Plenio, Phys Rev Lett. {\bf 105}, 190501 (2010).

\bibitem{andre11} A. R. R. Carvalho and M. F. Santos, New J. Phys. {\bf 13}, 013010 (2011).

\bibitem{santos11} M. F. Santos and A. R. R. Carvalho, Europhys. Lett. {\bf 94}, 64003 (2011).

\bibitem{santos11b} M. F. Santos, M. Terra Cunha, R. Chaves, and A. R. R. Carvalho, Phys. Rev. Lett. {\bf 108}, 170501 (2012).

\bibitem{bennett13} S. D. Bennett, N. Y. Yao, J. Otterbach, P. Zoller, P. Rabl, and M. D. Lukin, 
arXiv:1301.2968. 

\bibitem{arenz13} C. Arenz, C. Cormick, D. Vitali, and G. Morigi, arXiv:1303.1977.

\bibitem{bela} V. P. Belavkin, ``Nondemolition measurements and control in quantum dynamical systems'', 
in {\em Information Complexity and Control in Quantum Physics}, vol. {\bf 294}, p. 311 (Springer, New York, 1987).

\bibitem{gardinerq}C. Gardiner and P. Zoller, {\em Quantum Noise} (Springer, New York, 2010). 

\bibitem{wisebook}{H. M. Wiseman and G. J. Milburn, {\em Quantum measurement and control} 
(Cambridge University Press, New York, 2010).}

\bibitem{myrev} See A. Serafini, ISRN Optics, 2012, 275016 (2012), and references therein.


\bibitem{WisMilFeedback} 
H. M. Wiseman and G. J. Milburn,
Phys. Rev. Lett. {\bf 70}, 548 (1993); Phys. Rev. A {\bf 49}, 1350 (1994).

\bibitem{armen02}M. A. Armen, J. K. Au, J. K. Stockton, A. C. Doherty and H. Mabuchi,
\prl {\bf 89}, 133602 (2002).

\bibitem{wido05}{H. M. Wiseman and A. C. Doherty, \prl {\bf 94}, 070405 (2005).}

\bibitem{diosiwis}{H. M. Wiseman and L. Di\'osi, Chem. Phys. {\bf 268}, 91 (2001).}

\bibitem{mancini05}{S. Mancini and J. Wang,  Eur. Phys. J. D {\bf 32}, 257 (2005); 
S. Mancini, \pra {\bf 73}, 010304(R) (2006).}

\bibitem{mancini07}{S. Mancini and H. M. Wiseman, \pra {\bf 75}, 012330 (2007).}

\bibitem{seramancio}{A. Serafini and S. Mancini, \prl {\bf 104}, 220501 (2010).}

\bibitem{nurdin12}J. Nurdin and N. Yamamoto, Phys. Rev. A {\bf 86}, 022337 (2012).

\bibitem{xiao87} M. Xiao, L.-A. Wu, and H. J. Kimble, Phys. Rev. Lett. {\bf 59}, 278 (1987).

\bibitem{mon06} A. Monras, Phys. Rev. A {\bf 73}, 033821 (2006).

\bibitem{genoni11} {M. G. Genoni, S. Olivares and M. G. A. Paris, Phys. Rev. Lett. {\bf 106}, 
153603 (2011)}

\bibitem{braunvl} S. L. Braunstein and P. van Loock, Rev. Mod. Phys. {\bf 77}, 513 (2005).

\bibitem{gaucluster}N. C. Menicucci, S. T. Flammia, and O. Pfister, Phys. Rev. Lett. {\bf 101}, 130501 (2008); 
M. Ohliger, K. Kieling, and J. Eisert, Phys. Rev. A {\bf 82}, 042336 (2010).

\bibitem{furusawa}A. Furusawa, J. L. S{\o}rensen, S. L. Braunstein, C. A. Fuchs, H. J. Kimble, and E. S. Polzik, 
Science {\bf 282}, 706 (1998); H. Yonezawa, T. Aoki, and A. Furusawa, Nature {\bf 431}, 430 (2004).

\bibitem{sqqkd}M. D. Reid, Phys. Rev. A {\bf 62}, 062308 (2000);
D. Gottesman and J. Preskill, Phys. Rev. A {\bf 63}, 022309 (2001); 
R. Garcia-Patron and N. J. Cerf, Phys. Rev. Lett. {\bf 102}, 130501 (2009). 

\bibitem{ASM}{A. Ferraro, S. Olivares and M. G. A. Paris, {\em Gaussian States in Quantum
Information}, (Bibliopolis, Napoli, 2005).}

\bibitem{Hol75} A. S. Holevo, IEEE Trans. Inf. Theor. {\bf IT21} 533 (1975); 
R. Simon, N. Mukunda, and B. Dutta, \pra {\bf 49}, 1567 (1994).

\bibitem{note2}{Since we are only interested in stationary, steady-state situations, we will not have to worry with the dependence of $\varrho$ on $\Delta t$ or $t$, which will be hence safely neglected in what follows.}

\bibitem{passive}{M. M. Wolf, J. Eisert, and M. B. Plenio, \prl {\bf 90}, 047904 (2003).}

\bibitem{pirandolo}S. Pirandola, A Serafini, and S. Lloyd, \pra {\bf 79}, 052327 (2009).

\bibitem{bathia}$\forall$ $k$-dimensional subspace $\Sigma_{k}$ and 
hermitian $M$, $\exists$ $\ket{v}\in\Sigma_{k}$ 
such that $\bra{v}v\rangle=1$ and $\bra{v}M\ket{v}\le\ldn{k}(M)$. 
See, {\em e.g.}, 
R. Bathia, {\em Matrix Analysis} (Springer, New York, 1996), page 58.

\bibitem{vidal02} G. Vidal and R. F. Werner, Phys. Rev. A {\bf 65}, 032314 (2002).

\bibitem{plenioln}{M. B. Plenio, \prl {\bf 95}, 090503 (2005).}



\bibitem{note4}{As already emphasised in the seminal EPR paper, 
continuous variable squeezing and entanglement are intimately related, in that the latter implies the former and, conversely, the former can always be converted into the latter by rotations in the phase space.}

\bibitem{squash} B. C. Buchler, M. B. Gray, D. A. Shaddock, T. C. Ralph and D. E. McClelland,
Opt. Lett. {\bf 24}, 259 (1999);
H. M. Wiseman, J. Opt. B, Quant. and Semiclass. Opt.  \textbf{1}, 459 (1999);
S. Mancini, \emph{et al.}, J. Opt. B, Quant. and Semiclass. Opt.  \textbf{2}, 190 (2000).

\bibitem{tomcanali} T. Tufarelli, A. Retzker, M. B. Plenio and A. Serafini, 
New Journal of Physics {\bf 14}, 093046 (2012).

\bibitem{wisemanInLoop} H. M. Wiseman, Phys. Rev. Lett. {\bf 81}, 3840 (1998).

\bibitem{note1}{For system's temperature $T$ -- in natural units -- and 
bare mode frequencies $\{\omega_j\}$, one would have $N_j=1/(\exp(\omega_j/T)-1)$.}

\bibitem{kappa}  From now on we will set $\kappa=1$, and use the loss rate as a unit of frequency.
Notice that, to simplify our notation, we have assumed the same loss rate $\kappa$ affecting all modes. 


\bibitem{teraphonons} D. G. Cahill, F. Watanabe, A. Rockett, and C. B. Vining, Phys. Rev. B {\bf 71}, 235202 (2005). 

\bibitem{teraphotons} J. E. Schaar {\em et al.}, IEEE J. Sel. Top. Quantum Electr. {\bf 14}, 354 (2008).

\bibitem{reid89}M. D. Reid, \pra {\bf 40}, 913 (1989).

\bibitem{smith02}W. P. Smith, J. E. Reiner, L. A. Orozco, S. Kuhr and H. M. Wiseman, 
Phys. Rev. Lett. {\bf 89}, 133601 (2002).

\bibitem{sayrin12}C. Sayrin {\em et al.}, Nature {\bf 477}, 73 (2011).

\bibitem{gardiner} C. W. Gardiner, {\em Handbook of Stochastic Methods} (Springer, Berlin, 1985).






\end{thebibliography}
\end{document}